\documentclass[12pt]{article}
\usepackage{amssymb}
\usepackage{graphicx}

\renewcommand{\theequation}{\thesection.\arabic{equation}}

\newlength{\extraspace}
\newlength{\extraspaces}
\newcommand{\be}{\begin{equation}
\addtolength{\abovedisplayskip}{\extraspaces}
\addtolength{\belowdisplayskip}{\extraspaces}
\addtolength{\abovedisplayshortskip}{\extraspace}
\addtolength{\belowdisplayshortskip}{\extraspace}}
\newcommand{\ee}{\end{equation}}

\newcommand{\ba}{\begin{eqnarray}
\addtolength{\abovedisplayskip}{\extraspaces}
\addtolength{\belowdisplayskip}{\extraspaces}
\addtolength{\abovedisplayshortskip}{\extraspace}
\addtolength{\belowdisplayshortskip}{\extraspace}}
\newcommand{\ea}{\end{eqnarray}}

\newcommand{\newsection}[1]{
\vspace{12mm}
\pagebreak[3]
\addtocounter{section}{1}
\setcounter{equation}{0}
\setcounter{subsection}{0}
\setcounter{footnote}{0}
\noindent{\bf \thesection. #1}
\nopagebreak
\medskip
\nopagebreak}

\newcounter{saveeqn}
\newcommand{\alpheqn}{\setcounter{saveeqn}{\value{equation}}%
 \stepcounter{saveeqn}\setcounter{equation}{0}%
 \renewcommand{\theequation}
     {\thesection.\mbox{\arabic{saveeqn}\alph{equation}}}}
\newcommand{\reseteqn}{\setcounter{equation}{\value{saveeqn}}%
  \renewcommand{\theequation}{\thesection.\arabic{equation}}}

\begin{document}
\addtolength{\baselineskip}{1.5mm}

\thispagestyle{empty}
\begin{flushright}
\end{flushright}
\vbox{}
\vspace{1.5cm}

\begin{center}
{\LARGE{A coherent-state-based path integral for quantum mechanics on the Moyal plane
 }}\\[16mm]
{H.~S. Tan\footnote{Email:haisiong.tan@alumni.nus.edu.sg~~}}
\\[6mm]
{\it Raffles Junior College, Department of Physics, 10 Bishan Street 21, Singapore 574013}\\[2mm]
{\it }

\end{center}
\vspace{2cm}

\centerline{\bf Abstract}\bigskip
\noindent
Inspired by a recent work that proposes using coherent states to evaluate the Feynman kernel in noncommutative space,
we provide an independent formulation of the path-integral approach for quantum mechanics on the Moyal plane, with
the transition amplitude defined between two coherent states of mean position coordinates. In our approach, we invoke
solely a representation of the noncommutative algebra in terms of commutative variables. The kernel
expression for a general Hamiltonian was found to contain gaussian-like damping terms, and it is non-perturbative in the sense
that it does not reduce to the commutative
theory in the limit of vanishing $\theta$ - the noncommutative parameter.
As an example, we studied the free particle's propagator which turned out to be oscillating
with period being the product of its mass and $\theta$. Further,
it satisfies the Pauli equation for a charged particle with its spin aligned to a constant, orthogonal $B$ field
in the ordinary Landau problem, thus providing an interesting evidence of how noncommutativity can induce spin-like effects
at the quantum mechanical level.


\newpage
\newsection{Introduction}

When noncommutative geometry was first formally introduced by Snyder in \cite{Snyder}, it was presented as a possible strategy
to regulate the divergences of quantum field theories. By replacing the spacetime manifold by a noncommutative algebra represented in
a Hilbert space of states, the
notion of a spacetime point becomes a cell of which size is characterized by the noncommutative parameter $\theta$.
One can then hope to regularize the divergences in a similar spirit as using UV cut-off on momenta integrations. However, the
standard technique of using Weyl quantization and the Groenewold-Moyal star product \cite{Moyal} to obtain noncommutative quantum field theory
yields a new class of divergences known as UV/IR mixing \cite{Seiberg}. For example, in noncommutative $\phi^{4}$ theory in $4D$ \cite{Szabo},
a phase factor
$\textrm{e}^{ik_{i}\theta^{ij}p_{j}}$ where $k,p$ denote momenta, is present in the perturbative diagrams. This causes many previously
divergent terms to be generally convergent due to rapid oscillation at high momenta, but in the infrared limit $p\rightarrow 0$, the
effect of $\theta$ disappears and UV divergences are restored.

This problem currently plagues a large class of noncommutative field theories \cite{Szabo}, with merely a couple of surviving exceptions like the
Wess Zumino model \cite{Silva} and supersymmetric Yang-Mills theory with 16 supercharges \cite{YM}. It is a stringy feature that comes with
using the Groenewold-Moyal star product when quantizing fields in noncommutative spacetime. As is well known, the star product can
be shown in general cases to be equivalent to using the Bopp's shift \cite{Bopp} {\it i.e.} treating the noncommutative coordinates as a
linear combination of commutative coordinates and momenta such that the noncommutative algebra is preserved.

In the context of the path integration method, there have been interesting
expositions of ways of formulating it in noncommutative spacetime.  In general these methods attempt
to evaluate noncommutative analogues of the ordinary Feynman kernel -
\be
\mathcal{K}(x,t; x_{0},t_{0})=\langle x \vert \hat{U}(t,t_{0}) \vert x_{0} \rangle
\ee
where $\hat{U}$ is the unitary time evolution operator, and $\vert \, x \rangle \equiv \vert x_{1} x_{2} \dots x_{d} \rangle$ are the position eigenkets
in $d$ dimensions. Noncommutative geometry implies the absence of common position eigenstates. This problem was circumvented in,
for example, \cite{Dragovich}, where the
noncommutative kernels were obtained from the commutative ones by transforming phase space coordinates via the Bopp's shift.
Thus, the kernel is defined in terms of the auxiliary {\it commuting} variables but it is important to note that they are not physical.
On the other hand, in \cite{Acatrinei} and \cite{Vaquera}, this problem
is cleverly avoided by defining the kernel to be the transition amplitude
between two states with prescribed position along the first axis of coordinates and well defined momentum along the second axis -
in other words, doing phase space path integrals. However, one might be tempted to furnish a closer analogue to $(1.1)$ by defining the kernel
to be the transition amplitude between states of {\it mean} positions (since states of sharp position eigenvalues are inadmissible).
Such a formulation would be arguably more natural as it captures the fuzziness of a noncommutative space.

Interestingly, and as the primary source of inspiration for our work here, Smailagic and Spallucci formulated path integrals \cite{Smailagic_1}
by taking coherent states to define the kernel. As was also suggested in \cite{Bander}, the coherent states in this
context are eigenstates of complex combinations of the position operators and as we shall discuss in details later, they are states of definite
{\it mean} positions. Thus, they act as a meaningful set of replacement for the position eigenstates admissible only in the commutative theory.
What is notable in the approach in \cite{Smailagic_1} is that the free particle's propagator turns out to contain a damping exponential term.
This was argued to lead to a UV finite corresponding quantum field theory \cite{Smailagic_2} with divergence-less
loop diagrams - thus solving the UV/IR mixing pathology mentioned above.

We will construct a path-integral model on the noncommutative Moyal plane in this paper, following the trick in \cite{Smailagic_1} of
using coherent states in defining the fundamental Feynman kernel. However, as the reader is urged to compare, our derivation process,
the final expressions for the kernel and its resulting physics will be quite different. The paper is organized as follows: in Section 2,
we present our program after outlining some fundamental principles and explicitly work out the general expression for the
noncommutative kernel; in Section 3, we evaluate the path integral for a free particle and describe some implications
for its dynamics. The paper ends with a brief discussion of possible future work. We will use naturalised units where $\hbar = c = 1$.

\newsection{A coherent-state based path integral}

We begin with the commutator relations for the position $\hat{X}_{i}$ and momentum operators $\hat{P}_{i}, i = 1,2$ in the Moyal plane:
\alpheqn
\ba
\label{commutator}
&&[\hat{X}_{i},\hat{X}_{j}]=i\theta \epsilon_{ij} \,,\\
&&[\hat{P}_{i},\hat{P}_{j}]=0\,,\\
&&[\hat{X}_{i},\hat{P}_{j}]=i\delta_{ij}\,,
\ea
\reseteqn
where $\theta$ is the noncommutative parameter and $\epsilon_{ij}$ is the totally antisymmetric tensor of rank 2 with $\epsilon_{12}=1$.
In most conventional papers on noncommutative quantum mechanics (see, for examples, \cite{NCQM}), in particular
those concerned with phenomenology, one usually
proceeds by replacing the usual product between functions with the Groenewold-Moyal star-product. This is then equivalent to performing
the well-known Bopp's shift defined as $\tilde{X}_{i}\rightarrow X_{i} - \frac{1}{2}\theta \epsilon_{ij} P_{j}$, where the new coordinates
$\tilde{X}_{i}$ together with the momenta generate quantum mechanics on the usual commutative manifold. This convenient technique has been
used extensively in both quantum mechanics and quantum field theories built on noncommutative spaces \cite{Chaichian}.
In particular,  in \cite{Dragovich}, it
was proposed that by relating between the Lagrangians in the commutative and noncommutative regimes (via Bopp's shift), one can obtain
directly the noncommutative path integral. In such a model, it was found that there was no correction to the free particle,
while other quadratic potentials like the harmonic oscillator yield kernels which reduce continuously to the corresponding commutative cases.
Another point to take note is that field theories built on such approaches generally suffer from UV/IR divergences in their perturbative dynamics.

In this paper, we study another approach to path-integrals in noncommutative quantum mechanics. The guiding principle of our program
is simple - first, we define quantum states which contain information of both the noncommutative coordinates and are thus
eigenstates of a linear combination of these operators. It turns out, and as noted also in a number of papers \cite{{Sochichiu},{Bander}},
that they are none other than coherent states - very similar to those of the harmonic oscillator yet, as we shall point out later, different
in certain important aspects. The kernel is then the transition amplitude between the initial and the final coherent state
which has evolved in time according
to Schrodinger equation. The physical meaning, as we will also elaborate later, is that these coherent states represent states which
have definite {\it{mean}} position values. This constrasts interestingly with the normal approaches in literature which actually deal
with transition amplitudes between
position eigenstates which are {\it{not}} the physical coordinates but rather algebraic representations of the noncommutative coordinates. Of course,
the convenience in the latter approach lies in that one can control the limiting process from the noncommutive theory to its commutative one, but as
we shall observe later, dealing directly with physically meaningful noncommutative variables can bring us new surprises\footnote{This point was
also raised in \cite{Harikumar} in the context of noncommutative quantum mechanics with gauge potentials. It was noted that the usage of star product
yields gauge dependent answers, while working directly with noncommutative variables and the Seiberg Witten map seemed more appropriate.}.

As mentioned, the basic principle of our approach is not new. In \cite{Smailagic_1} and \cite{Smailagic_2},
Smailagic and Spallucci formulated the path integral
on a noncommutative plane using coherent states which are identical to the ones used here.
It was shown that the propagator for a free particle exhibits UV cut-off induced by the noncommutative
parameter $\theta$, because the propagator in momentum space was calculated to be of the form
$\frac{exp(-\theta p^{2} /2)}{p^{2}+m^{2}}$ \cite{Smailagic_1}
and the corresponding quantum field theory is then UV finite with divergence-less loop diagrams.

A crucial ingredient in their derivations was that the expectation value (taken with respect to the coherent states)
of the plane wave operator $\textrm{exp}(iP \cdot X)$
was defined as the
quantum wave function of a free point particle on the noncommutative plane. Herein lies the difference between theirs and our approach here. As will be
shown later, instead of this heuristic, we invoke solely
the algebraic representations of the noncommutative phase space (2.1) to derive the path integration.
It turns out that we will obtain very distinct results.

Now, let us proceed from (2.1) by representing the algebra of (2) on ${\it{L}}^{2}$-integrable functions of $x_{1},\,x_{2}$ via
\alpheqn
\ba
\label{rep}
&&\hat{X}_{i}  \mapsto x_{i} - \frac{\theta \epsilon_{ij}}{2} \,  \frac{1}{i} \frac{\partial}{\partial x_{j}} \,,\\
&&\hat{P}_{i}  \mapsto \frac{1}{i} \frac{\partial}{\partial x_{i}}\,,
\ea
\reseteqn
Further, we define the operators $A$, $A^{\dagger}$ as
\alpheqn
\ba
\label{annihilation}
&&A =  \big( \hat{X}_{1} + i\hat{X}_{2} \big)  \,,\\
&&A^{\dagger} = \big(  \hat{X}_{1} - i\hat{X}_{2} \big) \,,\\
&&[A, A^{\dagger}] = 2\theta \,,
\ea
\reseteqn
If we denote $\vert\alpha \rangle$ as the eigenstate of $A$, with $A \vert\alpha\rangle = \alpha \vert\alpha\rangle$,
then effectively, we have coherent states as our basis for the
phase space defined in (2.1), with (2.3c) being the Heisenberg-Weyl algebra up to a scaling factor $2\theta$.
Properties of coherent states are well-studied (see, for example, \cite{coherent}).
In the usual context of one-dimensional simple harmonic oscillators,
these coherent states minimize the Heisenberg uncertainty relation, with the real and imaginary parts of
the eigenvalues $\alpha$ being proportional to the mean position and momentum respectively. In contrast, our coherent
states are states of definite mean positions $X_{i}$ since we have
\alpheqn
\ba
\label{alpha}
&& \langle\alpha\vert X_{1} \vert\alpha\rangle =  \langle\alpha\vert \frac{A+A^{\dagger}}{2} \vert\alpha\rangle
= \textrm{Re} (\alpha) \langle \alpha \vert \alpha \rangle \equiv \bar{x}_{1\alpha} \langle \alpha \vert \alpha \rangle   \,,\\
&& \langle\alpha\vert X_{2} \vert\alpha\rangle =   \langle\alpha\vert \frac{A-A^{\dagger}}{2i} \vert\alpha\rangle
= \textrm{Im} (\alpha) \langle \alpha \vert \alpha \rangle \equiv \bar{x}_{2\alpha}\langle \alpha \vert \alpha \rangle  \,,
\ea
\reseteqn
To furnish the quantity $\langle x \vert \alpha \rangle$ where $x$ is the commutative coordinate
used in the representation (2.2), we have to solve the linear equation
\be
\label{alphax}
\Big( x_{1} + \frac{\,\theta}{2}\,\frac{\partial }{\partial x_{1}} +ix_{2} + i\frac{\,\theta}{2}\,\frac{\partial}{\partial x_{2}}
\Big) \langle x \vert \alpha \rangle
= \alpha \langle x \vert \alpha \rangle
\ee
where the LHS of (2.5) is just the representation of the operator $A$ following the prescription in (2.2).
Let $\langle x \vert \alpha \rangle = \textrm{e}^{u+iv}$ where $u, v$ are real functions of $x$. By solving for the real and imaginary parts of (2.5),
it is straightforward to show that its general solution is, up to a multiplicative constant,
\be
\langle x \vert \alpha \rangle = \textrm{exp} \Big( - \frac{1}{\theta} \Big( (x_{1}-\bar{x}_{1\alpha})^{2} + (x_{2}-\bar{x}_{2\alpha})^{2} \Big) +
V(-x_{2}, x_{1}) + iV(x_{1},x_{2}) \Big)
\ee
where $V(x)$ is any solution to the 2$D$ Laplace equation $\nabla^{2}V(x)=0$. Consider the $\alpha=0$ case in which we demand
\be
\lim_{\theta \rightarrow 0} \langle x \vert 0 \rangle = \delta^{2}(x)
\ee
since as $\theta$ vanishes, we want to recover the commutative theory as much as possible. Eqns (2.6) and (2.7) then yield an unique
renormalization constant for $\langle x \vert 0 \rangle$ as
\ba
\langle x \vert \alpha = 0 \rangle &=& \frac{1}{\pi \theta} \textrm{exp} \Big( -\frac{1}{\theta} ({x_{1}}^{2} + {x_{2}}^{2}) \Big)\,,\\
\langle 0 \vert 0 \rangle &=& \int\vert \langle  x \vert 0 \rangle \vert^{2} \textrm{d}^{2} x = \frac{1}{2\pi \theta}
\ea
Further, by choosing $V(x) = {\bar{x}}_{1\alpha} x_{2} - x_{1} {\bar{x}}_{2\alpha}$, we have
\be
\langle x \vert \alpha \rangle = \frac{1}{\pi \theta} \textrm{exp} \Big(\frac{-\vert \alpha \vert^{2}}{4\theta}\Big) \textrm{exp}
\bigg( -\frac{1}{\theta} \Big( (x_{1}-\frac{\alpha}{2})^{2} + (x_{2}+\frac{i\alpha}{2})^{2} \Big) \bigg)
\ee
which not only ensures $(2.8)$ but also the inner product between 2 coherent states to be
\ba
\langle \alpha \vert \beta \rangle &=& \int \int \langle \alpha \vert x \rangle \langle x \vert \beta \rangle \textrm{d}^{2}x\,\cr
&=&\frac{1}{2\pi \theta} \textrm{exp}\bigg( \frac{1}{2\theta} \Big( - \frac{\vert \alpha \vert^{2}}{2} -\frac{\vert \beta \vert^{2}}{2} + \beta \alpha^{*} \Big) \bigg)
\ea
Alternatively, following the conventional treatment of coherent state theory, define
\be
\vert \alpha \rangle = \textrm{exp} \bigg( \frac{\alpha A^{\dagger} - \alpha^{*} A}{2\theta} \bigg) \vert 0 \rangle
\ee
Then, together with $(2.9)$, we can invoke the standard Campbell-Baker-Hausdorff relations to arrive at $(2.11)$.
Consider now the limit of vanishing $\theta$. Although $(2.8)$ offers a continuous transition to the commutative theory, $(2.10)$
and $(2.11)$ become ill-defined in such a procedure. The wavefunction $\langle x \vert \alpha \rangle$ is a non-perturbative solution
and does not reduce smoothly to the commutative theory although we still enjoy
\be
\lim_{\theta \rightarrow 0} \vert \langle \alpha \vert \beta \rangle \vert = 2\delta^{2}( x_{\alpha} - x_{\beta} )
\ee
In this aspect, it is interesting to observe that in \cite{Smailagic_1}, the choice of $V(x)=0$ in (2.6) was effectively made, giving
nicely a smooth $\theta \rightarrow 0$ limit. However, it is straightforward to show that for such a choice, we would have, instead of (2.11)
\be
\langle \alpha \vert \beta \rangle_{V=0} = \frac{1}{2\pi\theta}\textrm{exp} \bigg( \frac{1}{2\theta} \Big( - \vert \alpha \vert^{2} - \vert \beta \vert^{2}
+ \beta \alpha^{*} + \beta^{*}\alpha \Big) \bigg)
\ee
We argue that it is important to prefer (2.10), in particular when we implement our path integration later by inserting sets of immediate coherent states,
because (2.10) yields
\be
\int \langle \alpha \vert \gamma \rangle \langle \gamma \vert \beta \rangle \textrm{d}^{2} \gamma = \langle \alpha \vert \beta \rangle
\ee
which is invalid for (2.14) even up to any multiplicative constant. Our choice of (2.10) and thus (2.11) implies that we have $\int \vert \alpha
\rangle \langle \alpha \vert \textrm{d}^{2} \alpha = 1$ with respect to the subspace of coherent states. This is a critical ingredient in carrying out
the path integration as follows: consider an initial state at time $t=t_{0}$,
denoted by $\vert\alpha\,\,t_{0}\rangle$ since $\alpha$ labels its mean position in the 2$D$
noncommutative plane.
We are interested in finding the transition amplitude $\langle\alpha^{'}\,\,\,t^{'}\,\vert\alpha\,\,\,t_{0}\rangle$ where $\vert\alpha^{'}\,\,t^{'}\rangle$
is the state ket which has evolved in time.
In our model, and as first proposed in \cite{Smailagic_1}, we define the noncommutative Feynman kernel as
$\langle\alpha^{'}\,\,t^{'}\,\vert\alpha\,\,\,t_{0}\rangle$. As in the ordinary case, we first split the time interval into $N$ equal
small slices with $t^{'} - t_{0} = N \varepsilon$, and insert complete sets of basis states
$\int \vert \alpha_{n}\,\,t_{n}\rangle \langle \alpha_{n} t_{n} \vert
\textrm{d}\alpha_{n}$ at each of the grid points $n = 1,2,\dots, N-1$. Thus, the fundamental entity of the integral is
\be
\label{kernel}
\langle\alpha^{'}\,\,t^{'}\,\vert\alpha\,\,\,t_{0}\rangle= \int \textrm{d}\alpha_{N-1} \dots \int \textrm{d}\alpha_{1}
\langle\alpha^{'}\,\,t^{'}\,\vert\alpha_{N-1}\,\,t_{N-1}\rangle    \langle\alpha_{N-1}\,\,t_{N-1}\,\vert\alpha_{N-2}\,\,\,t_{N-2}\rangle
\dots \langle\alpha_{1}\,\,t_{1}\,\vert\alpha\,\,\,t_{0}\rangle     \,.
\ee
where $\textrm{d}\alpha_{n} = \textrm{dRe}(\alpha_{n})\,
\textrm{dIm}(\alpha_{n})$. As usual, if we assume that the Dyson series remains valid,
then each matrix element can be approximated to first order in $\varepsilon$ as
\be
\label{step}
\langle\alpha_{n+1}\,\,t_{n+1}\,\vert\alpha_{n}\,\,t_{n}\rangle= \langle\alpha_{n+1}\vert 1 -
i\varepsilon \hat{H} \big(\hat{X}, \hat{P}\big) \vert   \alpha_{n}\rangle
+ O(\varepsilon^{2})
\ee
Since there is dependence of $\hat{H}$ on $\hat{P}$, we insert in a complete set of momentum eigenstates, thus
\ba
\label{momentum}
\int \textrm{d}p_{n} \textrm{d} p'_{n}   \langle\alpha_{n+1}\,\,\vert p_{n}\rangle   \langle p_{n}\,\,\vert 1 -
i\varepsilon \hat{H} \big(\hat{X}, \hat{P}\big) \vert
p'_{n}\rangle \langle p'_{n}\,\,\vert \alpha_{n}\rangle \cr
= \int \textrm{d} p_{n} \langle \alpha_{n+1} \vert p_{n} \rangle \langle p_{n} \vert \alpha_{n} \rangle
\Big(1 - i\varepsilon \hat{H} \big({\bar{\alpha}}_{n},
p_{n} \big) \Big) + O(\varepsilon^{2})\,,
\ea
Due to the choice of representation for $\hat{P}$ being equivalent
to its counterpart in the commutative quantum theory, it is straightforward that we have the familiar 2D plane wave:
\be
\langle x \vert p \rangle = \frac{1}{2\pi} \textrm{exp}\big(i(p_{1}x_{1}+p_{2}x_{2})\big)
\ee
Using (2.10) and (2.19), we can calculate $\langle \alpha \vert p \rangle$ explicitly as
\ba
\label{alphap}
\langle \alpha \vert p \rangle &=& \int \langle \alpha \vert x \rangle\langle x \vert p \rangle\, \textrm{d}x \cr
&=& \frac{1}{2\pi^{2}\theta}\, \textrm{exp}\big(-\frac{\vert \alpha \vert^{2}}{4\theta}  \int_{-\infty}^{\infty} \textrm{d}x_{1}
\int_{-\infty}^{\infty} \textrm{d}x_{2}\, \textrm{exp} \Bigg( ip\cdot x - \frac{1}{\theta} \Big( (x_{1} -  \frac{\alpha^{*}}{2}   )^{2} +  (x_{2}
- i \frac{\alpha^{*}}{2}   )^{2} \Big) \Bigg) \cr
&=& \frac{1}{2 \pi} \,\textrm{exp} \bigg(-\frac{\vert \alpha \vert^{2}}{4\theta}-\frac{\theta \vert p \vert^{2}}{4} + \frac{i}{2} \alpha^{*}
\Big(p_{1}+ip_{2}\Big)\bigg) \,,
\ea
We can now evaluate (2.17) and thus the entire path integral. Substituting (2.20) into (2.18),
\ba
\label{a1}
\langle \alpha_{n+1}\,\,t_{n+1} \vert \alpha_{n}\,\,t_{n} \rangle &&= \int \textrm{d}p_{n} \frac{1}{4\pi^{2}} \textrm{exp} \bigg(
-\frac{\vert \alpha_{n+1} \vert^{2}}{4\theta}-\frac{\theta \vert p_{n} \vert^{2}}{2} + \frac{i}{2} {\alpha_{n+1}}^{*}
\Big(p_{n1}+ip_{n2}\Big) \bigg) \times \cr
&&\textrm{exp} \bigg(
-\frac{\vert \alpha_{n} \vert^{2}}{4\theta} - \frac{i}{2} {\alpha_{n}}
\Big(p_{n1}-ip_{n2}\Big) \bigg) \times \bigg( 1 - i \varepsilon H\big( p_{n}, \bar{\alpha}_{n} \big) \bigg) \cr
&&=\frac{1}{4\pi^{2}} \textrm{exp} \bigg(-\frac{\vert \alpha_{n+1} \vert^{2} +\vert \alpha_{n} \vert^{2}}{4\theta} \bigg)
\int \textrm{d}p_{n}  \bigg( 1 - i \varepsilon H\big( p_{n}, \bar{\alpha}_{n} \big) \bigg) \times \cr
\textrm{exp} \bigg( - \frac{\theta \vert p_n \vert^{2}}{2}+&&\frac{i}{2}\varepsilon
(p_{n1}-ip_{n2}) \big( \frac{\alpha_{n+1} - \alpha_{n}}{\varepsilon}
\big) - \frac{1}{2} \Big( -2 \textrm{Im} (\alpha_{n+1}) p_{n1} + 2 \textrm{Re} (\alpha_{n+1})p_{n2} \Big) \bigg)\cr
&&
\ea
We now take the limit $\varepsilon \rightarrow 0$, $N \rightarrow \infty$ (while keeping $N\varepsilon = t^{'}-t_{0}$ constant)
to evaluate the kernel. Invoking the well-known representation of the exponential function:
\be
\lim_{N\rightarrow\infty} \Big( 1 + \frac{x}{N} \Big)^{N} = \textrm{e}^{x}
\ee
and by replacing the discrete quantities by continuous ones,
\be
\frac{\alpha_{n+1} - \alpha_{n}}{\varepsilon} \rightarrow \dot{\alpha} (t_{n})\,,\qquad \varepsilon \sum^{N-1}_{n=0} f(t_{n})
\rightarrow \int^{t^{'}}_{t_{0}} \textrm{d}\tau f(\tau)
\ee
we have from (2.21)
\ba
\label{a2}
\langle \alpha'\,\,t' \vert \alpha \,\,t_{0} \rangle &&= \lim_{N \rightarrow \infty} \Big( \frac{1}{4\pi^{2}}\Big)^{N}
\int \prod_{m=1}^{N-1} \textrm{d} \alpha_{m} \prod_{n=0}^{N-1} \textrm{d} p_{n} \textrm{exp} \Big(i\int_{t_{0}}^{t'}\textrm{d}\tau
\frac{1}{2}(p_{1}-ip_{2})\dot{\alpha} - H(p,\alpha) \Big) \times \cr
&&\textrm{exp}\bigg( - \frac{\vert \alpha_{n+1} \vert^{2} + \vert \alpha_{n} \vert^{2}+2\theta^{2} \vert p_{n} \vert^{2}}{4\theta} -
\Big[p_{n2} \textrm{Re} (\alpha_{n+1}) - p_{n1} \textrm{Im} (\alpha_{n+1}) \Big] \bigg)
\ea
To make the notation more concise, we can introduce complex momentum $\mathcal{P} = \frac{1}{2}(p_1+ip_2)$
to write (2.24) more elegantly as
\be
\langle \alpha'\,\,t' \vert \alpha \,\,t_{0} \rangle \sim \int \mathcal{D}\alpha \int \mathcal{D}\mathcal{P}\, \textrm{exp} \bigg(
- \frac{\vert \alpha \vert^{2}}{2\theta} - 2\theta\vert \mathcal{P} \vert^{2} - 2 [ \alpha \wedge \mathcal{P} ] +i\int \textrm{d} \tau \mathcal{P}^{*} \dot{\alpha} - H \bigg)
\ee
which is the final form of our noncommutative kernel.

\newsection{The Free Particle}

In this section, we will study the propagator for the free particle moving on the noncommutative plane starting from
(2.21). We assume the hamiltonian in this case to be
\be
\hat{H}_{\textrm{{\scriptsize free}}} = \frac{\hat{P}^{2}}{2m}
\ee
It turns out in our work that the simple form of (3.1) is actually quite deceptive because, as we shall observe after our calculation,
it implies some rather unexpected physics. After substituting (3.1) into (2.21), we integrate over all momenta $\mathcal{P}$ in the
entire $\mathbf{R}^{2}$ analytically, often using the Gaussian relation:
\be
\int_{-\infty}^{\infty} \textrm{d} Q \textrm{exp} \bigg( - i\varepsilon \Big(A (Q + \lambda)^{2} - \beta \Big) \bigg)
= \sqrt{\frac{\pi}{i\varepsilon A}} \textrm{exp}
\Big(i\varepsilon \beta \Big)
\ee
to evaluate the integral
\ba
&&\int^{\infty}_{-\infty} \textrm{d}p_{n1} \textrm{d}p_{n2} \bigg( \textrm{exp} \Big( - \frac{\theta}{2} \vert p_{n} \vert^{2} + \frac{i}{2}(p_{n1}-ip_{n2})
(\alpha_{n+1} - \alpha_{n})+ \textrm{Im} (\alpha_{n+1})p_{n1} - \textrm{Re}(\alpha_{n+1})p_{n2}-i\varepsilon \frac{p^{2}_{n}}{2m}\Big)\bigg)\,\cr
&&=\frac{2m\pi}{m\theta+i\varepsilon} \textrm{exp} \bigg( \alpha_{n} \alpha^{*}_{n+1} \frac{m}{2(m\theta + i\varepsilon)} \bigg)
\ea
The path-integral is then reduced to be one involving only $\alpha$ integration and using (3.3) we obtain
\be
\label{a3}
\langle \alpha'\,\,t' \vert \alpha \,\,t_{0} \rangle = \lim_{N \rightarrow \infty}  \Big(\frac{\beta}{2\pi\theta}\Big)^{N}
\int \prod_{i=1}^{N-1} \textrm{d}^{2} \alpha_{i}    \textrm{exp} \Big( -\frac{1}{4\theta}\Big(\vert \alpha_{i+1} \vert^{2}  + \vert \alpha_{i} \vert^{2}
-2\beta \alpha_{i} {\alpha^{*}}_{i+1}             \Big)
\ee
where $\beta \equiv 1 - \frac{i\varepsilon}{m\theta + i\varepsilon}$. (Note that we have still kept $\varepsilon$ so that as $N\rightarrow\infty$,
we obtain the correct limit for (3.4).) Now consider the product of an arbitrary series of $j$, $j+1$, $j+2$ terms in (3.4)
and define $\alpha \equiv U + iV$. It can be shown after tedious algebra that for any $\gamma \epsilon \mathbf{C}$,
\ba
&&\Big( -\frac{\vert\alpha_{j}\vert^2 + \vert\alpha_{j+1}\vert^2   }{2} + \gamma \alpha_{j} \alpha^{*}_{j+1} \Big) +
\Big( -\frac{\vert\alpha_{j+1}\vert^2 + \vert\alpha_{j+2}\vert^2   }{2} + \beta \alpha_{j+1} \alpha^{*}_{j+2} \Big)\,\cr
&&\,\cr
&=&\Big( -\frac{\vert\alpha_{j+2}\vert^2 + \vert\alpha_{j}\vert^2   }{2} + \gamma \beta \alpha_{j} \alpha^{*}_{j+2} \Big)
- \Big(U_{j+1} + \frac{\beta}{2}F_{j,j+2} \Big)^{2} - \Big(V_{j+1} + \frac{\beta}{2}G_{j,j+2} \Big)^{2}
\ea
where $\beta F= \gamma \alpha_{j} + \beta \alpha^{*}_{j+2}$, $\beta G= -i ( \gamma \alpha_{j} - \beta \alpha^{*}_{j+2})$. Thus, each integration
step over $\textrm{d}\alpha_{j+1} = \textrm{d}U_{j+1} \textrm{d}V_{j+1}$ in (3.5) would yield a factor of $\pi$ and increase the power of $\beta$
by 1. This reduction formula leads to
\be
\langle \alpha'\,\,t' \vert \alpha \,\,t_{0} \rangle = \lim_{N \rightarrow \infty} \frac{\beta^{N}}{2\pi\theta}
\textrm{exp} \Big( -\frac{1}{4\theta}\Big(   \vert \alpha' \vert^{2} + \vert \alpha \vert^{2} + 2\beta^{N} \alpha \alpha^{'*}\Big) \Big)
\ee
Finally, since (2.22) implies $\lim_{N\rightarrow\infty\,,\,\varepsilon\rightarrow0} \beta^{N} = \textrm{exp} \bigg(\frac{-i(t^{'}-t_{0})}{m\theta} \bigg)$ , we have
the final form of the free particle's propagator as
\be
\langle \alpha'\,\,t' \vert \alpha \,\,t_{0} \rangle = \frac{1}{2\pi\theta}\textrm{exp} \Big(\frac{-i(t^{'}-t_{0})}{m\theta}\Big)
\textrm{exp} \bigg( -\frac{1}{4\theta} \Big( \vert \alpha' \vert^{2} + \vert \alpha \vert^{2} + 2\textrm{exp}\Big(\frac{-i(t^{'}-t_{0})}{m\theta}\Big) \alpha \alpha^{'*} \bigg)
\ee
An immediate consistency check is to take the limit $t^{'}-t_{0} \rightarrow 0$ of (3.7) which recovers comfortably (2.11) - the expected transition amplitude
between two coherent states. An interesting novel feature in (3.7) is that the propagator can be described
to be {\it spinning} around the origin,
with the period characterized by $m\theta$. We would observe later that indeed, its effective dynamics involves
angular momentum in an interesting
manner. Also, as already observed in (2.10), the wavefunction $\langle x \vert \alpha \rangle$ and thus (3.7) does not have a well-defined $\theta\rightarrow0$
limit.

To elucidate phenomenological consequences further, let us make the physical interpretation as was decided in (2.4) -
\be
\alpha =\bar{x}_{1} + i\bar{x}_{2}
\ee
where $(\bar{x}_{1},\bar{x}_{2})$ describe the mean positions of a free particle of mass $m$ on the
noncommutative plane as measured by the observer.
In the commutative theory, the free particle's propagator is the Green's function of the Schrodinger equation.
It turns out, after some deliberate manipulation, that our noncommutative kernel (in the coordinates $\bar{x}_{i}$) satisfies the equation
\be
\bigg[ -\frac{1}{2m} \bar{\nabla}^{2} + \frac{\omega}{2i}( \bar{x}_{2} \partial_{\bar{x}_{1}} - \bar{x}_{1} \partial_{\bar{x}_{2}})
+ \frac{m\omega^{2}\bar{x}^{2}}{8} + \frac{\omega}{2} - i \frac{\partial}{\partial t} \bigg] \mathcal{K}(\bar{x},t; \bar{x}_{0},t_{0}) = 0
\ee
where $\omega = \frac{1}{m\theta}$. The hamiltonian operator in (3.9) can be written more suggestively as
\ba
\hat{H}_{eff}(\bar{x}) &=& \frac{1}{2m}\Big( \frac{1}{i}\frac{\partial}{\partial \bar{x}_{i}} - q\mathcal{A}(\bar{x}) \Big)^{2} +\frac{\omega}{2}\,\\
\mathcal{A} &=& \frac{1}{2}(B\hat{z} \times \vec{\bar{x}})\,,\,\,\,\,\,\, qB=-m\omega=-1/\theta\,
\ea
Thus, the kernel of our path integral model implies that, as described by its mean observed positions,
the particle is confined on the plane by a constant magnetic field $B = -1/q\theta$ in the orthogonal direction
induced by noncommutativity, with $q$ as its charge. Also, the extra energy term $\frac{\omega}{2} = -\frac{qB}{2m}$ turns out to be
equivalent to the interaction energy of the spin of a spin-$1/2$ particle with the magnetic field.  Indeed, the effective hamiltonian operator
is just a component of the Pauli hamiltonian for the Landau problem in symmetric gauge \cite{Harikumar}
\ba
\hat{H}_{Pauli}(\bar{x})&=& \frac{1}{2m}\Bigg[ \Big(\frac{1}{i}\frac{\partial}{\partial \bar{x}_{i}} - q\mathcal{A}(\bar{x}) \Big)^{2}
+ \frac{g}{2}qB\sigma_{3} \Bigg]\,\cr
&=& \left( \begin{array}{ccc} \hat{H}_{eff}(g) & 0 \\ 0 & \hat{H}_{eff} (-g) \end{array} \right)
\ea
where $\vec{\sigma}$ are the Pauli matrices, and $g=2$ is the gyromagnetic ratio. To recall, we have begun with a spinless free particle
on the Moyal plane, yet surprisingly, the propagator in our framework turns out to describe physics of a charged, spin-$1/2$ particle
confined to the plane by a $B$ field to which the spin is aligned.

Actually, it is not difficult to discern evidence of non-commutative geometry in the physics of a charged particle in a magnetic field, and
thus to partially understand (3.9). Quantum mechanically (see for example, \cite{Ballentine}), such a particle's equation of motion follows an
ensemble of circular orbits of which centers $x_0, y_0$ obey the operator commutator
\be
[x_0, y_0] = \frac{i}{-qB} = \frac{i}{m\omega}
\ee
where the cyclotron frequency $\omega$ bears the same definition in (3.11). Further, if the particle has spin, the spin vector precesses
about the $B$-field orthogonal to the plane.

On the other hand, the spin-magnetic coupling term coincides with the ground state energy of the 1$D$ harmonic oscillator. Indeed, if one compares (3.7)
with the coherent-state-path-integral for the 1$D$ harmonic oscillator which Klauder found in \cite{Klauder} to be
\be
\langle \alpha'\,\,t' \vert \alpha \,\,t_{0} \rangle = \textrm{exp} \Big(\frac{-i\omega(t^{'}-t_{0})}{2}\Big)
\textrm{exp} \bigg( -\frac{1}{2} \Big( \vert \alpha' \vert^{2} + \vert \alpha \vert^{2} + 2\textrm{exp}\Big(-i\omega(t^{'}-t_{0}) \Big) \alpha \alpha^{'*} \bigg)
\ee
it is easy to observe a striking similarity. Naively, the spin-interaction energy term $\omega/2$ appears to be related to the
identical term in the 1D harmonic oscillator hamiltonian. But to be careful, not only are the hamiltonians for (3.14) and (3.7) different, the
coherent states which we have defined here have distinct physical meanings from the standard ones used in (3.14).

This brings questions as to whether (i)higher-spin interaction
energy terms will fall out naturally in higher dimensional
generalisation of our framework, (ii)a more complete path-integral
with spin degrees of freedom will reveal spin precession in induced
magnetic fields, and (iii)how conventional coherent state path
integrals for harmonic oscillators are related precisely to our
model. It would be interesting to uplift our model along these
stated directions to uncover fully this surprising hint of a
relationship between noncommutative geometry and spin structures.
Indeed, in this aspect, a first step was made in \cite{Colatto} and
\cite{Djemai} where the noncommutative parameter $\theta$ is tied to
a local spin structure $S=m^{2}\theta$. In contrast, our progagator
has a spin-$1/2$ Pauli Hamiltonian partially emerging as the
effective theory.

\newsection{Future Directions}

We have constructed a path integral model for quantum mechanics on the noncommutative plane, and evaluated the propagator for the free particle
as a simple application. The fundamental kernel is defined as the transition amplitude between two coherent states at different times.
Interestingly, the free particle's propagator
satisfies the Pauli equation for a charged particle with its spin aligned to a constant, orthogonal $B$ field
in the ordinary Landau problem.
This result is distinct from previously known works in noncommutative path integration, primarily because
we have taken coherent states as the starting point of our derivation. These states are eigenvectors of the operator $A$ as defined in (2.3), and
describe physical states of mean coordinates $x_{1}, x_{2}$. Thus, they have the closest physical meaning to the commutative counterpart of
simultaneous eigenstates of the coordinate operators. Our solution is also non-perturbative in the sense that the propagator does not deform
continuously to the commutative theory in the vanishing $\theta$ limit. An immediate generalisation of our work would be to study the path-integral for other types of
interaction Hamiltonians to see how the results differ from those presented in \cite{Dragovich} where the Bopp's shift is used as
a mapping tool to obtain the noncommutative kernels from the commutative ones for quadratic Hamiltonians.

In \cite{Bander}, noncommutative quantum mechanics has been formulated on the premise that measurement of position operators, or functions
of such operators is determined by their expectation values between generalized coherent states \cite{perel} based on the group $SO(N,1)$.
This leads to $N$
dimensional rotation invariance. What is also interesting is that it provides an avenue for noncommutativity on other types of spaces -
including compact
ones - to be realised quantum mechanically via coherent states. For example, one may be able to analyse the quantum mechanics of
the noncommutative fuzzy sphere via generalised coherent states of $SU(2)$ following \cite{Bander}.
Thus, a possible further extension of our work would
be to perform coherent state-based path integration in higher dimensions and on certain topologies,
for example, on hyperspheres and compare results with the commutative versions.

Our approach has been inspired by \cite{Smailagic_1} in which the path integral was also formulated on the Moyal plane using coherent states. But as
mentioned earlier, our derivation has been very much different and so are the results. For the model in \cite{Smailagic_1} and \cite{Smailagic_2},
it was argued that it leads to UV finite quantum field theories due to the presence of Gaussian factors in the kernel providing exponential cut-off
for large momenta. It would be interesting to check similarly if we can enjoy a divergence-free quantum field theory via our approach
and thus avoid the problem
of UV/IR mixing\footnote{Recently in \cite{Grosse}, H. Grosse and R. Wulkenhaar gave solution of this problem for the special case of a scalar 4$D$ theory
defined on the Moyal-deformed space. The IR/UV mixing was taken into account by adding an oscillator term to the free Lagrangian, with the
model satisfying the Langmann-Szabo duality.}. Indeed, if we follow the spirit of argument in \cite{Smailagic_1}, it is likely that we would be able to achieve likewise due to the
Gaussian terms in (2.25) and (3.7). This may imply that
a subtle redefinition of the Feynman kernel yields noncommutative quantum field theories without the need for renormalisation.

Finally, as mentioned in Section 3, it would be interesting to generalise our framework to construct
path-integrals with spin degrees of freedom, in order to clarify and expand on the hint we have found here for possible linkages
between noncommutative geometry and spin structures in quantum theory.
\bigbreak\bigskip\noindent{{\bf Acknowledgement}}

\noindent The author would like to thank R. Parwani and the anonymous referees for many useful advice and suggestions.


\bigskip\bigskip
{\renewcommand{\Large}{\normalsize}
}
\end{document}